\documentclass[a4paper,11pt]{article}
\pdfoutput=1 

\usepackage{jinstpub} 

\usepackage[version=3]{mhchem}
\usepackage{siunitx}
\usepackage{natbib}
\usepackage{amsmath} 

\usepackage{verbatim}
\usepackage{multirow}
\usepackage{xcolor}

\usepackage{setspace}
\usepackage{graphicx}
\usepackage{amsmath}
\usepackage{siunitx}

\usepackage{pbox}
\usepackage{xcolor}

\newcolumntype{"}{@{\hskip\tabcolsep\vrule width 1pt\hskip\tabcolsep}}
\DeclareSIUnit[number-unit-product = \;]\year{yr}
\DeclareSIUnit\parsec{pc}
\DeclareSIUnit\torr{Torr}
\DeclareSIUnit\centimeter{\centi \meter}
\DeclareSIUnit\sq{\ensuremath{\Box}}
\usepackage{textcomp}
\usepackage{caption}
\usepackage{subcaption}
\usepackage{url}
\usepackage{ctable}

\usepackage[english]{babel}
\usepackage{hyperref}
\addto\extrasenglish{
    
}
\usepackage{lineno}

\title{Charge Amplification in Low Pressure CF$_4$:SF$_6$:He Mixtures with a Multi-Mesh ThGEM for Directional Dark Matter Searches}

\author[1]{F.D. Amaro,}
\author[4,5]{E. Baracchini,}
\author[6]{L. Benussi,}
\author[6]{S. Bianco,}
\author[7,8]{F. Borra,}
\author[6]{C. Capoccia,}
\author[6,9]{M. Caponero,}
\author[10]{D. S. Cardoso,}
\author[7,8]{G. Cavoto,}
\author[6]{I.A. Costa,}
\author[11]{T. Crane,}
\author[6]{E. Dané,}
\author[4,5]{M. D'Astolfo,}
\author[6]{G. Dho,}
\author[4,5]{F. Di Giambattista,}
\author[7]{G. D'Imperio,}
\author[7]{E. Di Marco,}
\author[1]{J.M.F. Dos Santos,}

\author[12]{A.C. Ezeribe,}

\author[4,5]{D. Fiorina,}
\author[7]{F. Iacoangeli,}
\author[8]{H.P. Lima Júnior,}
\author[14]{G.S.P. Lopes,}
\author[6]{G. Maccarrone,}
\author[1]{R.D.P. Mano,}
\author[15]{R.R. Marcelo Gregorio,}
\author[4,5]{D.J.G. Marques,}
\author[6]{G. Mazzitelli,}

\author[\color{blue}*\color{black},12]{A.G. McLean,\note[\color{blue}*]{\color{blue}Corresponding author.}}

\author[1]{C.M.B. Monteiro,}
\author[14]{R.A. Nobrega,}
\author[14]{I.F. Pains,}
\author[6]{E. Paoletti,}
\author[6]{L. Passamonti,}
\author[7,8]{S. Piacentini,}
\author[6]{D. Piccolo,}
\author[6]{D. Pierluigi,}
\author[7]{D. Pinci,}
\author[4,5]{A. Prajapati,}
\author[7]{F. Renga,}
\author[1]{R.J.d.C. Roque,}
\author[6]{F. Rosatelli,}
\author[6]{A. Russo,}
\author[6,16]{G. Saviano,}
\author[12]{A. Scarff,}
\author[12]{N.J.C. Spooner,}
\author[6]{R. Tesauro,}
\author[6]{S. Tomassini,}
\author[4,5]{and S. Torelli}

\affiliation[1]{LIBPhys, Department of Physics, University of Coimbra, 3004-516 Coimbra, Portugal}
\affiliation[2]{Istituto Nazionale di Fisica Nucleare, Sezione di Roma TRE, 00146 Rome, Italy}
\affiliation[3]{ Dipartimento di Matematica e Fisica, Università Roma TRE, 00146 Rome, Italy}
\affiliation[4]{Gran Sasso Science Institute, 67100 L’Aquila, Italy}
\affiliation[5]{Istituto Nazionale di Fisica Nucleare, Laboratori Nazionali del Gran Sasso, 67100 Assergi, Italy}
\affiliation[6]{Istituto Nazionale di Fisica Nucleare, Laboratori Nazionali di Frascati, 00044 Frascati, Italy}
\affiliation[7]{Istituto Nazionale di Fisica Nucleare, Sezione di Roma, 00185 Rome, Italy}
\affiliation[8]{Dipartimento di Fisica, Sapienza Università di Roma, 00185 Rome, Italy}
\affiliation[9]{ENEA Centro Ricerche Frascati, 00044 Frascati, Italy}
\affiliation[10]{Centro Brasileiro de Pesquisas Físicas, Rio de Janeiro, RJ 22290-180, Brazil}
\affiliation[11]{AWE plc, Aldermaston, Reading, Berkshire, RG7 4PR, United Kingdom}
\affiliation[12]{Department of Physics and Astronomy, University of Sheffield, South Yorkshire, S3 7RH, United Kingdom}
\affiliation[13]{Universidade Estadual de Campinas - UNICAMP, Campinas 13083-859, SP, Brazil}
\affiliation[14]{Universidade Federal de Juiz de Fora, Faculdade de Engenharia, Juiz de Fora, MG 36036-900, Brazil}
\affiliation[15]{School of Physical and Chemical Sciences, Queen Mary University of London, E1 4NS, United Kingdom}
\affiliation[16]{Dipartimento di Ingegneria Chimica, Materiali e Ambiente, Sapienza Università di Roma, 00185 Rome, Italy}

\emailAdd{ali.mclean@sheffield.ac.uk}

\abstract{The CYGNO collaboration is developing next generation directional Dark Matter (DM) detection experiments, using gaseous Time Projection Chambers (TPCs), as a robust method for identifying Weakly Interacting Massive Particles (WIMPs) below the Neutrino Fog. SF$_6$ is potentially ideal for this since it provides a high fluorine content, enhancing sensitivity to spin-dependent interactions and, as a Negative Ion Drift (NID) gas, reduces charge diffusion leading to improved positional resolution. CF$_4$, although not a NID gas, has also been identified as a favourable gas target as it provides a scintillation signal which can be used for a complimentary light/charge readout approach. These gases can operate at low pressures to elongate Nuclear Recoil (NR) tracks and facilitate directional measurements. In principle, He could be added to low pressure SF$_6$/CF$_4$ without significant detriment to the length of $^{16}$S, $^{12}$C, and $^{19}$F recoils. This would improve the target mass, sensitivity to lower WIMP masses, and offer the possibility of atmospheric operation; potentially reducing the cost of a containment vessel. In this article, we present gas gain and energy resolution measurements, taken with a Multi-Mesh Thick Gaseous Electron Multiplier (MMThGEM), in low pressure SF$_6$ and CF$_4$:SF$_6$ mixtures following the addition of He. We find that the CF$_4$:SF$_6$:He mixtures tested were able to produce gas gains on the order of 10$^4$ up to a total pressure of 100 Torr. These results demonstrate an order of magnitude improvement \cite{Miyomoto2004} in charge amplification in NID gas mixtures with a He component.}


\begin{document}
\maketitle
\raggedbottom

\section{Introduction}
\label{sec:intro}

There is overwhelming evidence for the existence of Dark Matter (DM) \cite{DARKMATTER}, constituting $\sim$ 85\% of the mass in the Universe \cite{planck}. A population of Weakly Interacting Massive Particles (WIMPs), hypothesised to have been created in the early Universe, provides a possible explanation. Many direct detection experiments have tried to measure rare elastic scattering events between WIMPs and nucleons and the sensitivity of these experiments has improved by several orders of magnitude since their inception \cite{Cushman}. Notable here are two-phase noble liquid Time Projection Chamber (TPC) DM experiments, like LZ and XENON \cite{LZ,XENON}. Their sensitivity is approaching the Neutrino Fog \cite{OHare2021}, the parameter space in which background neutrinos induce WIMP-like interactions. Current results show no excess in events near the Neutrino Fog \cite{LZ,XENON}, indicating the need for future searches to probe below this parameter space. In this eventuality, two-phase noble liquid TPCs will likely fail to effectively discriminate between neutrino and potential WIMP induced Nuclear Recoil (NR) signals \cite{neutrino,OHare2021,Billard2014}.

One method which could be exploited to search below the Neutrino Fog concerns the directional measurement of the recoiling nucleus \cite{direction}. This method enables discrimination between neutrino signals, predominantly originating from the Sun, and WIMP signals, which would appear to come from the Cygnus constellation due to the motion of the Solar System through the Galaxy. This Galactic signature would change direction over the course of a sidereal day, depending on the detector's latitude on Earth, further evidencing its Galactic origin \cite{Morgan2003}. Compared to an annual modulation signal, caused by Earth's motion around the Sun \cite{Spergel1988}, an anisotropic directional Galactic signature would be much more difficult to attribute to terrestrial phenomena \cite{Klinger2015,Kudryavtsev2010,Davis2014}. However, evidence shows such a measurement is not possible with current two-phase noble liquid TPC detectors \cite{DarkSide-20kCollaboration}.

The DRIFT experiments have pioneered an alternative directional search method using low pressure gaseous back-to-back Multi-Wire Proportional Counters (MWPCs) filled with a Negative Ion Drift (NID) gas mixture of CS$_2$:CF$_4$:O$_2$ at a total pressure of 41 Torr (30:10:1 Torr) \cite{Battat2017}. The NID gas CS$_2$ greatly enhanced DRIFT's position resolution by reducing diffusion in the drift region compared to electron drift gases \cite{Martoff2000}. However, generating sizable avalanches in NID gases is more challenging.  The NID gas SF$_6$ is now preferential to CS$_2$ due to its fluorine content, which is predicted to improve spin-dependant cross sections with a possible WIMP candidate, and because it is non-toxic and non-flammable \cite{Ellis1991, Cannoni2011, Gondolo, Phan2017}.  

As indicated in DRIFT above, directional detectors have been operated at low pressure. The low pressure operation allows the extension of the NR ionisation track to the mm-scale to better match the position resolution of the available charge readout technology. This allows the principle axis of a recoiling nucleus to be determined by reconstructing the track of ionised charge left behind in the gas. Measuring the relative charge density at either end of the track, the so-called head-tail effect, can be used to determine the direction of the recoiling nucleus along that principle axis \cite{Bat2016}. A detector capable of measuring the principle axis and head-tail effect in three dimensions can reduce the number of events needed for the positive identification of WIMPs to \(\mathcal{O}(10)\) \cite{Morgan2005}.   

Clearly the low pressure requirement here has implications for the detector volume necessary to achieve a given sensitivity, something that can not be mitigated by increasing the pressure since this would reduce directional sensitivity. However, an alternative possibility is to add He. The low density of this gas implies no significant impact on NRs from the heavier target gases. The addition of He would improve the detectors target mass, sensitivity to lower mass WIMPs, and offer the potential of atmospheric operation which would reduce the cost of a containment vessel \cite{Vahsen2020}. The demonstration of significant charge amplification in NID gas mixtures containing SF$_6$ and He is therefore highly desirable.






CF$_4$:He mixtures are the primary gas mixtures utilised by the CYGNO collaboration as these mixtures can provide a complimentary scintillation and charge readout approach. Although CF$_4$ is not a NID gas, this property can be obtained using a small admixture of SF$_6$ \cite{LaflerThesis}. It is therefore of interest to demonstrate significant charge amplification in CF$_4$:SF$_6$:He mixtures. A CF$_4$:SF$_6$:He gas mixture at nearly atmospheric pressure has been the subject of investigation by the CYGNO collaboration before \cite{Baracchini2018}, however the explicit measurement of the gas gain in such a NID mixture is yet to be realised. 


As mentioned, one challenge of using NID gases is that their electronegative nature can result in reduced avalanche gains. This comes from the requirement to first strip the electron before charge amplification can take place \cite{Phan2017}. Previous work with a Thick Gaseous Electron Multiplier (THGEM) showed that a single amplification stage was not sufficient for significant charge amplification in SF$_6$ at pressures < 100 Torr \cite{Callum_thesis}. It has been shown elsewhere that Gaseous Electron Multipliers (GEMs) and THGEMs typically require double or even triple GEM configurations to produce substantial gas gains in NID gases at both low and close to atmospheric pressure \cite{Phan2017, Baracchini2018, Ishiura2020}. Therefore, a multi-stage charge amplification device is likely required for successful NID detector operation. Recent results with a multi-stage Multi-Mesh ThGEM (MMThGEM) have demonstrated large (sub 10$^5$) gas gains in low pressure pure SF$_6$ \cite{McLean2023}, which makes the device a promising candidate for testing CF$_4$:SF$_6$:He mixtures for the next generation of directionally sensitive DM detectors. 

In this article, gaseous avalanche gain and energy resolution measurements, using the multi-stage MMThGEM, with low pressure pure SF$_6$ and a CF$_4$:SF$_6$ gas mixture are presented. Following these initial measurements, He was gradually added to the mixtures, and the effect on gas gain and energy resolution is presented.




\section{Experimental Apparatus and Gas Mixing Procedure}
\label{sec:design}

\begin{figure}[h]
    \captionsetup[subfigure]{justification=centering}
    \centering
    \begin{subfigure}{0.5\textwidth}
        \centering
        \includegraphics[width = 0.7\textwidth]{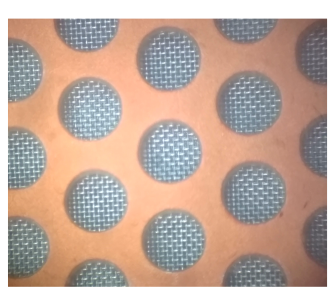}
        \caption{}
        \label{fig:MMThGEM_image}
    \end{subfigure}%
    \begin{subfigure}{0.5\textwidth}
        \centering
        \includegraphics[width = \linewidth]{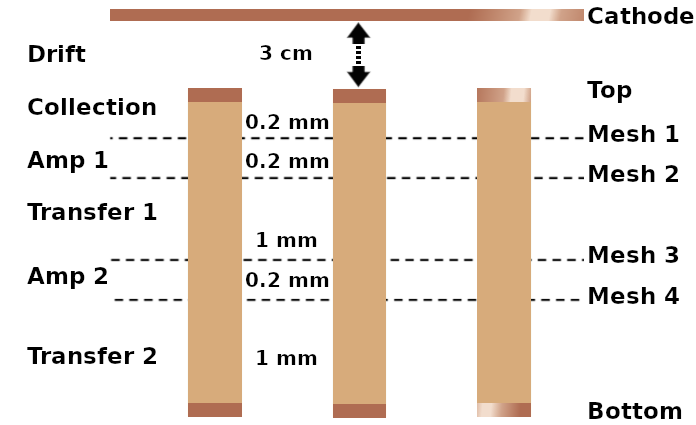}
        \caption{}
        \label{fig:MMThGEM_cross_sec}
    \end{subfigure}
    \caption{(a) Image of the MMThGEM hole structure as viewed from above. (b) Cross sectional diagram of the MMThGEM device/TPC. }
    \label{fig:MMThGEM_both}
\end{figure}

All studies undertaken in this work made use of a novel MMThGEM device developed by the authors in collaboration with CERN \cite{Olivera2018, Callum_thesis}. The MMThGEM is a multi-stage charge amplification device, similar in design to a regular THGEM, with the addition of intermediate mesh electrode layers seen to span across the holes in \autoref{fig:MMThGEM_image}. The holes have a diameter of 0.8 mm and a pitch of 1.2 mm.  As seen in \autoref{fig:MMThGEM_cross_sec}, the electrode layers designated as top, meshes 1-4, and bottom divide the detecter into six distinct regions when mounted together with a cathode above the MMThGEM. These are the drift, collection, amplification 1, transfer 1, amplification 2 and transfer 2 regions. Each electrode layer can be biased individually with High Voltage (HV) supplies to set up electric fields of varying strengths in the different regions. The device has a total thickness of 2.6 mm and a total active area of 10 x 10 cm. Further details of the device and operation can be found in Ref. \cite{Callum_thesis,Callum2023,McLean2023}.

When operated with a NID gas, the detector is expected to perform in the following way. Following an ionising event in the target volume between the cathode and top layer of the MMThGEM, the initial ionisation electrons will bind to the gas molecules to form Negative Ions (NIs). The drift field transports the NIs towards the MMThGEM. The collection field between the top and mesh 1 layers then guides the charge into the holes. The high field strength of the first amplification field accelerates the charge and energetic collisions cause the electron to strip from the NI and an avalanche of ionisation ensues. Evidence of stripping and avalanching charge in the device has been found previously with fields greater than 19000 V/cm \cite{Callum_thesis}.  When the avalanche electrons reach the first transfer field, the lower field strength means that the electrons can bind again to gas molecules to form NIs. The NIs are transported towards the second amplification field. Once the NIs reach the second amplification field the electron is stripped again and a second avalanche occurs. When coupled to a micromegas, which is its intended use, the second transfer field transports the amplified charge towards the micromegas readout plane for x-y positional measurements. However, for the purpose of the measurements presented here, the micromegas is not present and the amplified charge is measured on mesh 4 immediately after the second amplification field. As mentioned, this device has been tested previously and, following an optimisation procedure, was found to produce significant charge amplification of $\sim$ 9 x 10$^4$ in 40 Torr of SF$_{6}$  \cite{McLean2023}. This makes the device a suitable candidate for investigating charge amplification in SF$_6$:He mixtures.

Gain calculation was enabled by an $^{55}$Fe X-ray source positioned next to the MMThGEM TPC directed towards the centre of the drift volume. An average of $\approx$ 173 electron-ion pairs are produced following the photoelectric absorption of such an X-ray in SF$_6$ \cite{Lopes}. Test pulses injected into a CR-111 charge sensitive preamplifier and CR-200-4$\mu$s shaper connected to mesh 4 were used to determine the amount of charge reaching mesh 4, thus providing the gain calibration. During measurements, the TPC was biased by setting the cathode to a negative HV, the top electrode was grounded, and meshes 1 to 4 were positively biased. The current draw on the electrodes was used only to monitor sparking by tripping of the HV current limiter. Energy spectra were recorded on mesh 4 over the course of 30 minutes. A Gaussian function was fitted to the photo-peak observed in the energy spectra and the gain and energy resolution were ascertained from the mean and FWHM divided by the mean of the Gaussian fit function respectively. Error bars were determined from the uncertainties associated with the fitting procedure. The TPC setup, positioning of the $^{55}$Fe source, and method of gain calculation is identical to that presented in Ref. \cite{McLean2023}. 

\begin{figure}[b]
    \includegraphics[width=\textwidth]{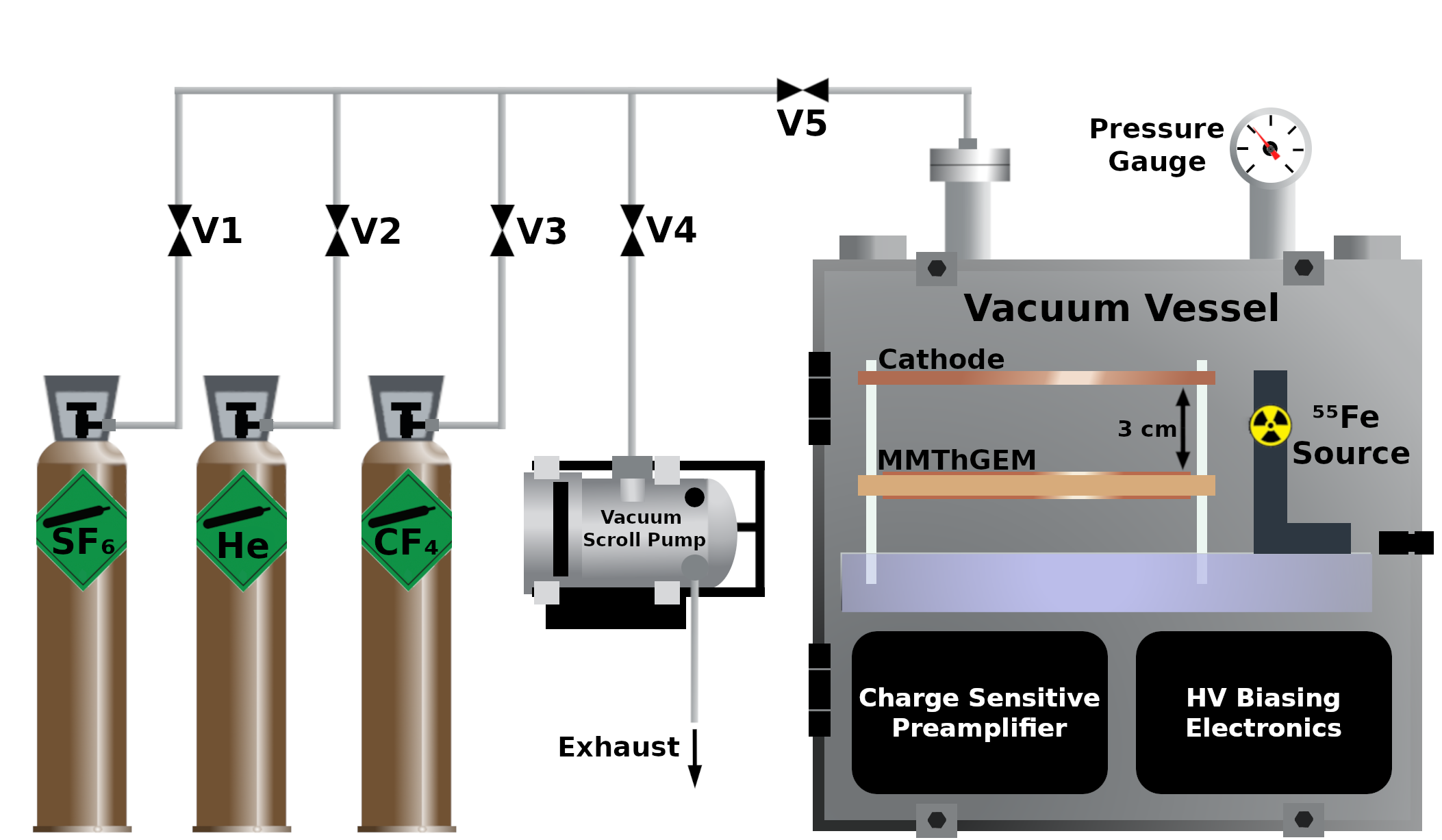}\caption{Diagram of the experimental setup and gas system used to fill the vacuum vessel with the desired mixture.}
    \label{fig:exp_diag}
    \centering
\end{figure} 

The diagram in \autoref{fig:exp_diag} illustrates the gas system used to fill the vacuum vessel. The filling procedure began by evacuating the vessel; with all gas bottles closed, valves V1-5 open, and the vacuum scroll pump turned on. The vessel was evacuated for a minimum of 48 hours achieving a vacuum < 10$^{-2}$ Torr. Following vessel evacuation, all valves were closed and the gas bottles briefly opened to reduce the pressure differential between the gas line and the lab. To begin filling, the SF$_6$ gas bottle and V1 were opened. Then V5 throttled the gas line to fill the vessel to the desired pressure. Then the SF$_6$ gas bottle, V1 and V5 were all closed. When a mixture was required, the gas line was first evacuated by opening V4 and turning on the pump between each addition. After 10 minutes of evacuating the gas line, V4 was closed and the pump was turned off again. Depending on whether He or CF$_4$ was required, either the He gas bottle and V2 or the CF$_4$ bottle and V3 were opened respectively. V5 was again used to fill the vessel to the desired partial pressure. Once the desired gas mixture was achieved, all valves and gas bottles were closed and gas gain measurements were taken with the MMThGEM TPC. Following each successful run of measurements the vessel was returned to the evacuation phase.


\section{Low Pressure Pure SF$_6$}
\label{sec:PureSF6}

\begin{figure}[h]
    \includegraphics[trim={3.5cm 0cm 3.9cm 1cm},clip,width=\textwidth]{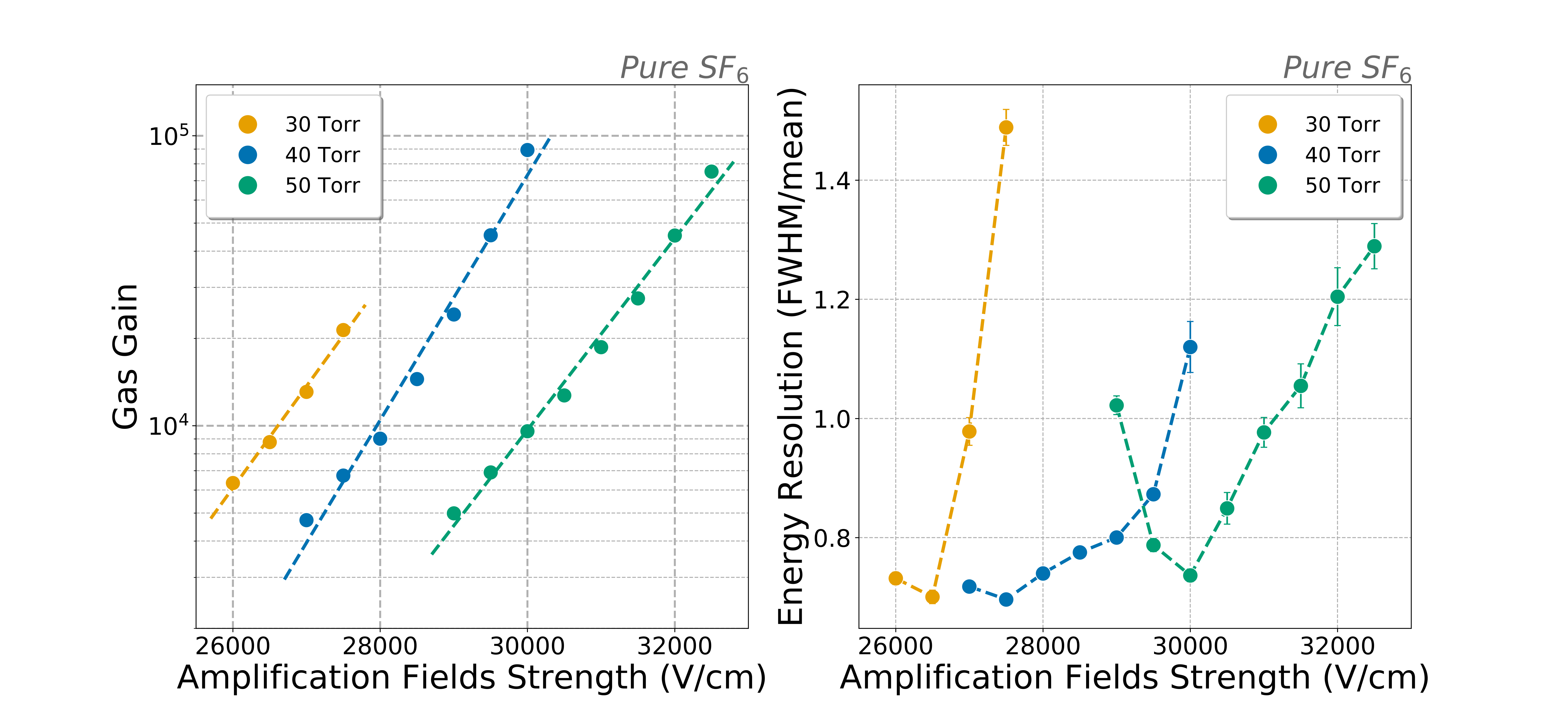}\caption{Effective gas gain vs amplification fields strength in 30, 40, and 50 Torr of pure SF$_6$ (left). Energy resolution vs amplification fields strength in 30, 40, and 50 Torr of pure SF$_6$ (right). Some error bars are smaller than the marker size and are therefore omitted from the graphs.}
    \label{fig:SF6_gain_30_40_50}
    \centering
\end{figure}

To begin these measurements, the vessel was filled with pure SF$_6$ to a pressure of 30 and 50 Torr by following the filling procedure described in \autoref{sec:design}. This was done in order to establish a set of baseline measurements for a small range of pressures before He was introduced; as discussed, measurements have already been taken with the MMThGEM in 40 Torr of SF$_6$. The cathode voltage, mesh 1 voltage, and transfer 1 field were set constant at -500 V, 40 V, and 900 V/cm based on the previous optimisation \cite{McLean2023}. The amplification fields were increased in tandem until an $^{55}$Fe photopeak could be resolved above the trigger threshold in the energy spectrum. The amplification fields were then increased in increments of 500 V/cm until sparking was observed on mesh 4. The result of these gain and energy resolution measurements, including previous measurements in 40 Torr \cite{McLean2023}, can be seen in \autoref{fig:SF6_gain_30_40_50}.

It is observed in \autoref{fig:SF6_gain_30_40_50} (left) that the gas gain increases exponentially with increasing amplification fields strength; dashed lines represent an exponential curve fitted to the data. The gain curves are seen to shift to the right as the gas pressure increases and larger electric fields are required to produce comparable gas gains. 


The maximum gas gains achieved before sparking occurred, \(G_{max}\), are summarised in \autoref{table:pureSF6}. Sparking and continuous ringing was observed in 30, 40, and 50 Torr at field strengths of 28000 V/cm, 30500 V/cm, and 33000 V/cm respectively; the observed ringing phenomena is discussed in more detail in Ref. \cite{Callum2023}. It is interesting to note that the maximum stable gas gain at 30 Torr is smaller than both 40 and 50 Torr. This was caused by earlier onset continuous ringing during the 30 Torr ramp up. Additionally, self-regulating ringing events which were able to return to baseline without intervention were noted to be more frequent in the 30 Torr run than the 40 and 50 Torr runs. This suggests that the ringing effect could be suppressed at higher pressures. A minority of these self-regulating events could also be responsible for the slight deviation from the exponential trend lines fitted to the 40 and 50 Torr data during the highest stable field strength exposures. 

\begin{table} [h!]
\centering
\captionsetup{justification=centering}
    \caption{Summary of pure SF$_6$ results including maximum stable gas gain, \(G_{max}\), and the minimum/maximum energy resolution, \(ER_{min}\) and \(ER_{max}\).}
    \label{table:pureSF6}
    \begin{tabular}{|c c c c|} 
     \hline
     Pressure (Torr) & \(G_{max}\times 10 ^{4}\) & \(ER_{min}\) & \(ER_{max}\) \\ [0.2ex] 
     \hline
     30 & 2.32 $\pm$ 0.04 & 0.701 $\pm$ 0.002  & 1.49 $\pm$ 0.03 \\ 
     40 & 8.93 $\pm$ 0.06  & 0.697 $\pm$ 0.001  & 1.12 $\pm$ 0.04 \\
     50 & 7.53 $\pm$ 0.05  & 0.737 $\pm$ 0.002 & 1.29 $\pm$ 0.04 \\ 
     \hline
    \end{tabular}

\end{table}



Previous measurements in 30 Torr of pure SF$_6$ with the MMThGEM were only capable of producing a maximum stable gas gain around $\sim$ 3000 before sparking occurred \cite{Callum2023}. This notable improvement is likely due to the optimisation procedure conducted with the MMThGEM \cite{McLean2023} and the lower drift field strength used since those initial measurements were taken. The lower drift field used here, 167 V/cm compared to 385 V/cm, likely reduces the charge density in the holes of the MMThGEM and therefore reduces the possibility of arcing between neighboring electrodes. 


The corresponding energy resolution curves are shown in \autoref{fig:SF6_gain_30_40_50} (right). This shows that the energy resolution initially decreases with increasing amplification field strengths before increasing significantly. The degrading energy resolution with increasing amplification fields strength could be caused by inefficient transfer of charge between the different regions, combined with the disruptive stripping/recombination of electrons. This is likely dictated by the changing field ratios with the neighbouring collection and transfer fields. The minimum and maximum energy resolutions, \(ER_{min}\) and \(ER_{max}\), achieved during the measurements are summarised in \autoref{table:pureSF6}. 40 Torr of SF$_6$ was subsequently selected as the base gas pressure for the following additions of He due to its superior gas gain and comparable energy resolution range \cite{McLean2023}.


\section{Sub-atmospheric SF$_6$:He Mixtures}
\label{sec:SF6Helium}

\begin{figure}[h]
    \includegraphics[trim={3.5cm 0cm 3.9cm 1cm},clip,width=\textwidth]{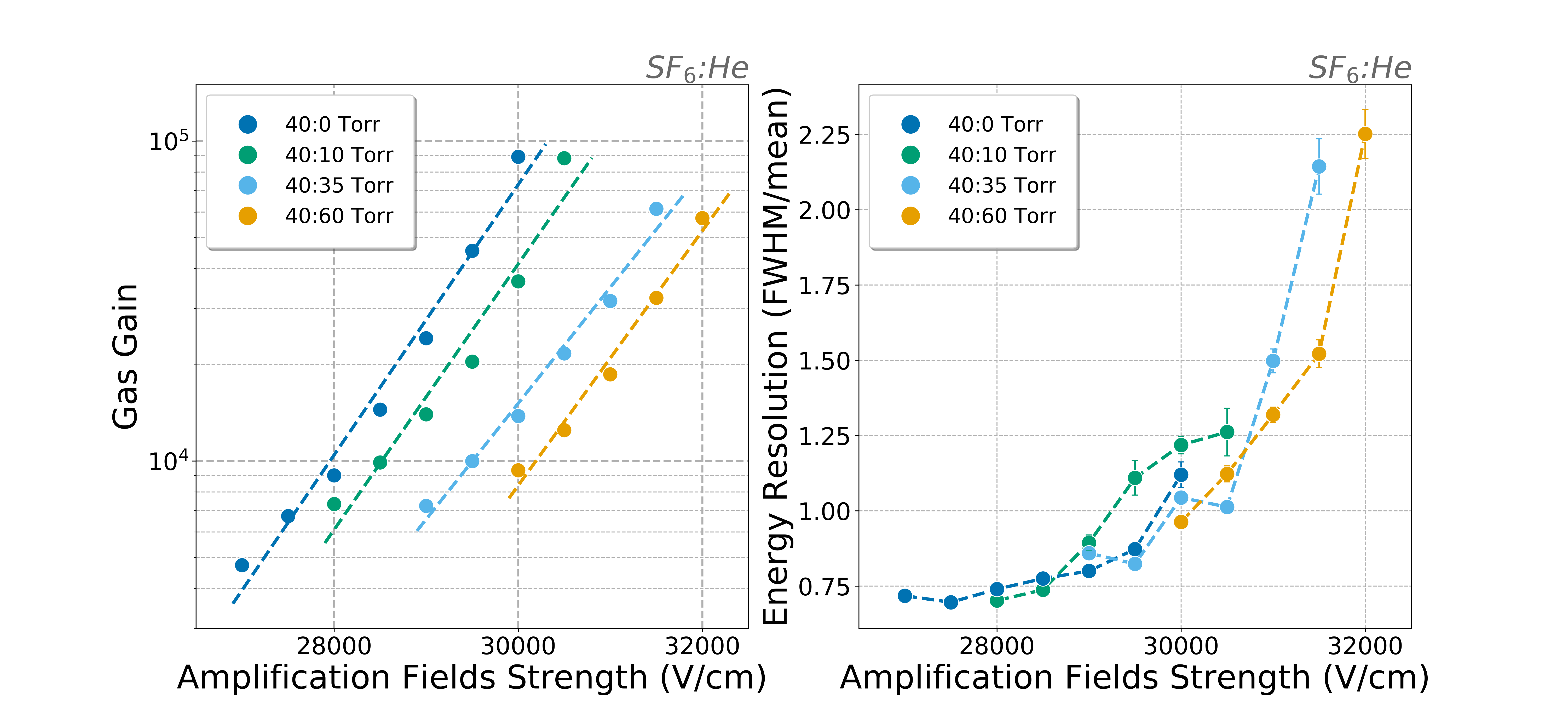}\caption{Effective gas gain vs amplification fields strength in SF$_6$:He mixtures (left). Energy resolution vs amplification fields strength in SF$_6$:He mixtures (right). Some error bars are smaller than the marker size and are therefore omitted from the graph.}
    \label{fig:SF6_Helium_gain}
    \centering
\end{figure}

Once the optimum base pressure of pure SF$_6$ was determined, see \autoref{sec:PureSF6}, the vessel was once again evacuated and filled with 40 Torr of SF$_6$. Helium was added to the vessel to bring the total pressure up to 50, 75, 100, 150, 380, and 760 Torr. These pressures were chosen so that a dynamic range of low and sub-atmospheric pressures could be tested. For each gas mixture the drift, collection, and transfer fields were set to the optimised settings \cite{McLean2023} and the amplification fields were again increased in tandem by increments of 500 V/cm. Spectra were recorded for stable operating voltages in each gas mixture and subsequent gas gains and energy resolutions were calculated. The results are presented alongside the 40 Torr pure SF$_6$ data in \autoref{fig:SF6_Helium_gain}.

As shown in \autoref{fig:SF6_Helium_gain} (left), not all mixtures tested were able to produce measurable gas gains; this is because some mixtures did not yield a distinct $^{55}$Fe photopeak before sparking occurred. However, the mixtures up to a total pressure of 100 Torr all exhibit exponential gas gain curves which shift to the right with increasing partial pressure of He. The maximum gas gains achieved for all SF$_6$:He mixtures are summarised in \autoref{table:SF6He}. Initially, a small addition of He does not appear to have any significant effect on the maximum gas gain, as the 40:10 mixture produces a similar maximum gas gain to pure SF$_6$. As more He is added to the vessel, the maximum stable gas gain begins to drop for the 
40:35 and 40:60 mixtures. This suggests that as more He is added to the vessel the maximum stable gas gain could drop further.

\begin{table}[h]
    \centering
    \captionsetup{justification=centering}
        \caption{Summary of SF$_6$:He results including maximum stable gas gain, \(G_{max}\), and the minimum/maximum energy resolution, \(ER_{min}\) and \(ER_{max}\).}
        \label{table:SF6He}
        \begin{tabular}{|c c c c|} 
         \hline
         SF$_6$:He Pressure (Torr) & \(G_{max} \times 10^4\) & \(ER_{min}\) & \(ER_{max}\) \\ [0.2ex] 
         \hline
         40:10 & 8.8 $\pm$ 0.2 & 0.70 $\pm$ 0.01 &  1.26 $\pm$ 0.08\\ 
         40:35 & 6.1 $\pm$ 0.2 & 0.82 $\pm$ 0.01 &  2.14 $\pm$ 0.09 \\
         40:60 &  5.7 $\pm$ 0.1 &  0.96 $\pm$ 0.01 & 2.25 $\pm$  0.08\\ 

         \hline
        \end{tabular}
\end{table}

The energy resolution measurements are also presented in \autoref{fig:SF6_Helium_gain} (right). The minimum and maximum energy resolutions are summarised in \autoref{table:SF6He}. It can be seen that the energy resolution worsens significantly with increasing partial pressure of He. For example, both SF$_6$:He mixtures with ratios 40:35 and 40:60 Torr produced fractional energy resolutions > 2. This results in the lower end of the photopeak merging with low level noise and therefore becomes more difficult to distinguish. It is therefore unsurprising that gas gains could not be determined at higher pressures, 150 - 760 Torr, when more He was added to the vessel. Any observed variation in the shape of these energy resolution curves is likely subject to the changing gas composition, pressure, and field ratios between the collection/transfer field and the amplification regions. 


These results demonstrate large gas gains, on the order of 10$^4$, in a low pressure NID mixture containing He up to a total pressure of at least 100 Torr for the first time. This constitutes an order of magnitude improvement on what has previously been achieved with CS$_2$:He mixtures \cite{Miyomoto2004}. Furthermore, the small addition of He will improve the sensitivity of the target gas to low WIMP masses in the context of a directional DM search.

\section{Low Pressure CF$_4$:SF$_6$ Mixture Optimisation}
\label{sec:SF6CF4Optimisation}

\begin{figure}[h]
    \includegraphics[trim={3.5cm 0cm 3.9cm 1cm},clip,width=\textwidth]{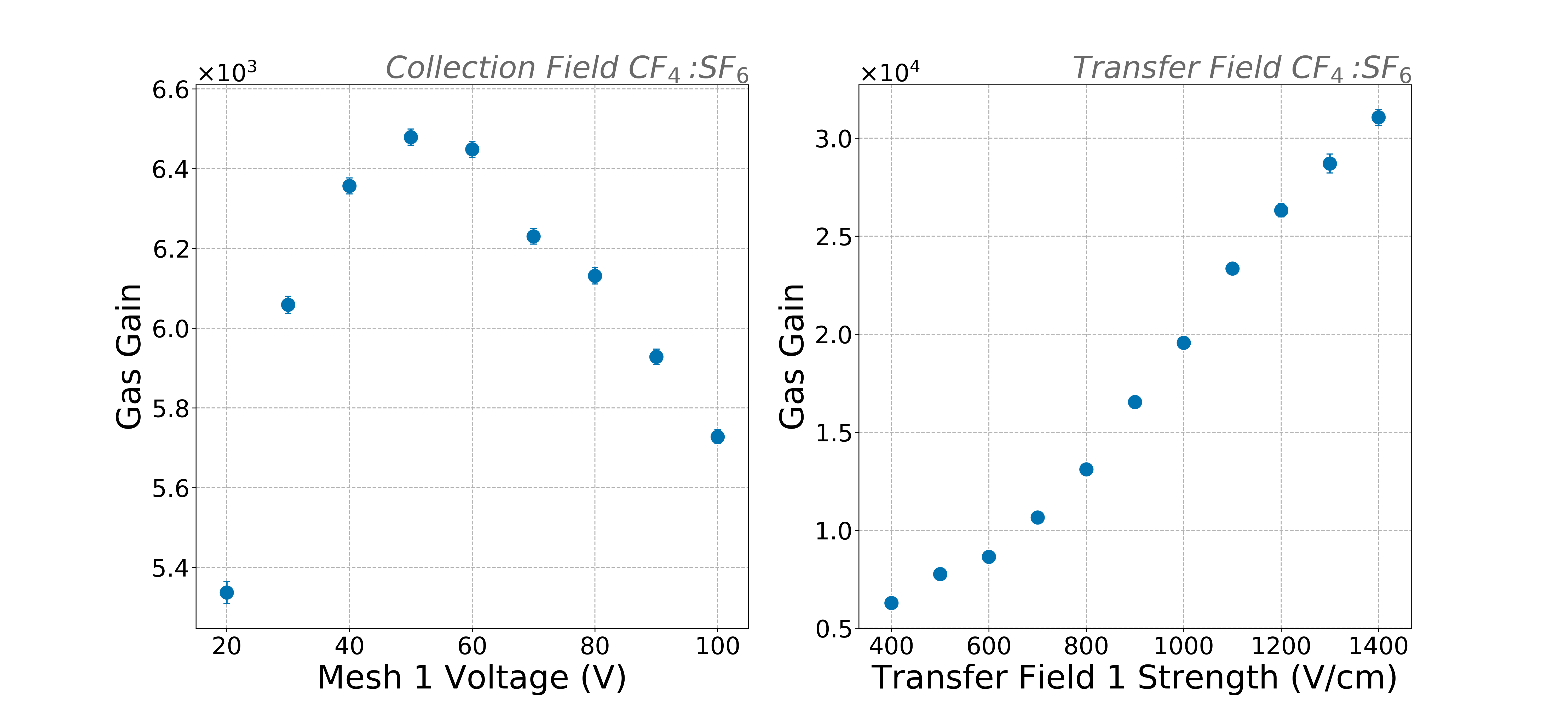}\caption{Effective gas gain vs mesh 1 voltage in a CF$_4$:SF$_6$ mixture with partial pressures 38:2 Torr (left). Effective gas gain vs transfer field 1 strength in a CF$_4$:SF$_6$ mixture with partial pressures 38:2 Torr (right). The majority of error bars are smaller than the marker size and are therefore not observed.}
    \label{fig:SF6_CF4_Collection}
    \centering
\end{figure}

As discussed in \autoref{sec:intro}, the CYGNO collaboration is interested in CF$_4$:SF$_6$:He mixtures due to the potential for a complimentary light/charge NID readout approach. Before introducing He, a 38:2 Torr base mixture of CF$_4$:SF$_6$ was subjected to an optimisation procedure identical to that presented in Ref. \cite{McLean2023}. Starting with the isolation of the collection field, the cathode voltage, amplification fields and transfer field 1 were held constant at -500 V, 25000 V/cm and 500 V/cm respectively. The collection field was then varied in isolation by increasing the mesh 1 voltage from 20 V to 100 V in increments of 10 V, the results of which are presented in \autoref{fig:SF6_CF4_Collection} (left). 

\autoref{fig:SF6_CF4_Collection} (left) shows that, as the mesh 1 voltage increases from 20 to 50 V the gas gain increases from 5340 \(\pm\) 30 to 6480 \(\pm\) 20. As the voltage increases further to 100 V the gas gain decreases to 5730 \(\pm\) 20. This trend is similar to what was observed in pure SF$_6$ however the peak occurs at 50 V rather than 40 V. This suggests that the optimisation could depend strongly on the gas mixture. As a clear peak is observed at 50 V, this voltage was taken as the optimum collection field for the CF$_4$:SF$_6$ base gas mixture. 


The transfer field was optimised in a similar way by holding the cathode voltage, mesh 1 voltage, and amplification fields constant at -500 V, 50 V and 25000 V/cm respectively. The transfer field was then varied in isolation from 400 V/cm to 1400 V/cm in increments of 100 V/cm and gain measurements were made for each transfer field strength. Results are shown in \autoref{fig:SF6_CF4_Collection} (right).


As the transfer field increases from 400 to 600 V/cm the gas gain increases gradually from 6290 \(\pm\) 40 to 8640 \(\pm\) 20. Between 600 and 1100 V/cm the gas gain increases more rapidly to 23400 \(\pm\) 100. Above 1200 V/cm the rate at which gas gain improves begins to slow up to 1400 V/cm producing a gas gain of 31100 \(\pm\) 400 before sparking was observed at 1500 V/cm. This trend is different to what was observed in 40 Torr of pure SF$_6$ which demonstrated a plateau above a field strength of 900 V/cm. As no plateau can be observed, the optimum transfer field was chosen to be 1400 V/cm as this produced the largest gas gain before sparking ocurred. 


\section{Sub-atmospheric CF$_4$:SF$_6$:He Mixtures}
\label{SF6CF4Helium}

\begin{figure}[h]
    \includegraphics[trim={3.5cm 0cm 3.9cm 1cm},clip,width=\textwidth]{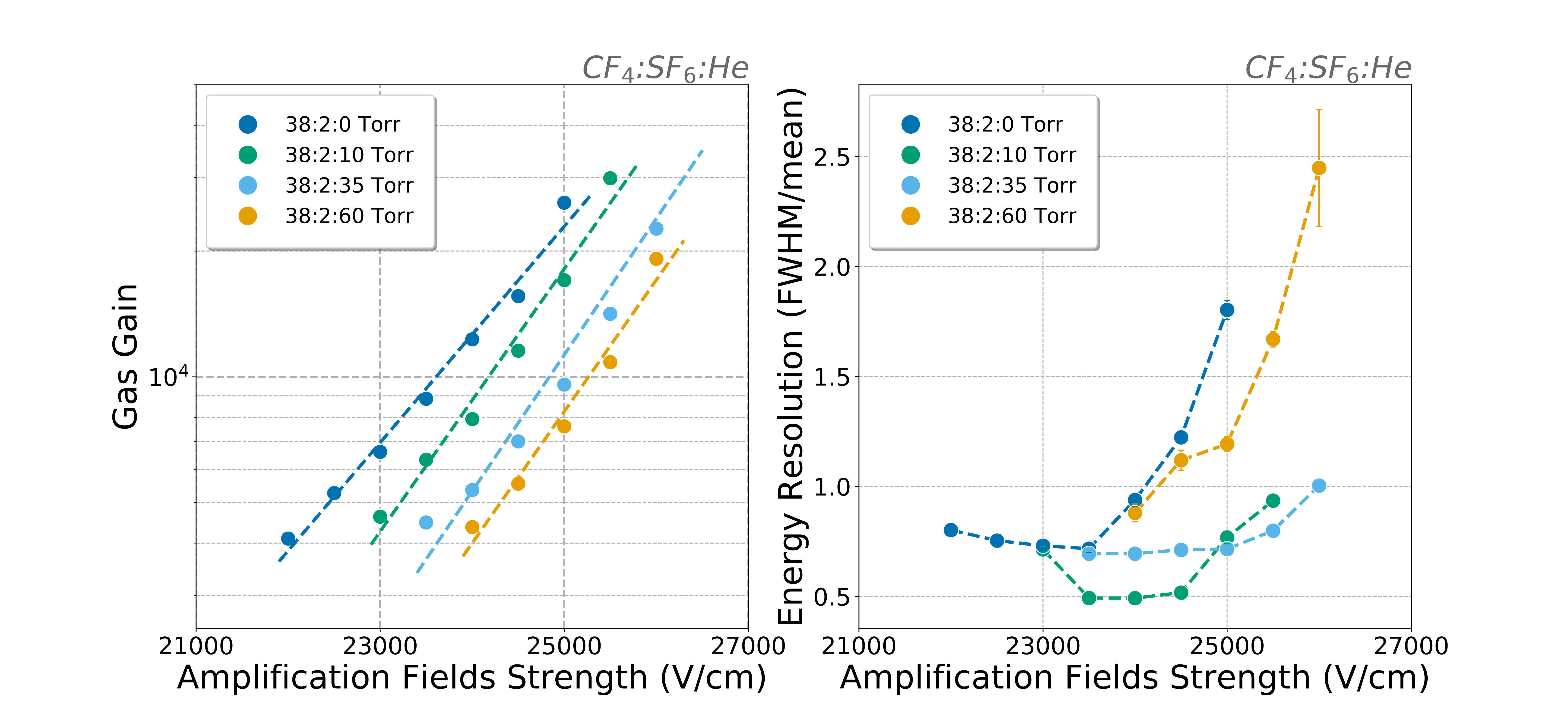}\caption{Effective gas gain vs amplification fields strength in CF$_4$:SF$_6$:He mixtures (left). Energy resolution vs amplification fields strength in CF$_4$:SF$_6$:He mixtures (right). Some error bars are smaller
    than the marker size and are therefore omitted from the graph.}
    \label{fig:SF6_CF4_He}
    \centering
\end{figure}

Following the gain optimisation of the collection and transfer fields in the MMThGEM for operation with the CF$_4$:SF$_6$ base mixture, He was again added to the vessel. Measurements were taken for total pressures of 50, 75, 100, 150, 380, and 760 Torr. The cathode voltage, mesh 1 voltage and transfer field were set to -500 V, 50 V and 1400 V/cm respectively. Then the amplification fields were again increased in increments of 500 V/cm and gain measurements were taken for each mixture until sparking ocurred. The results of this are presented in \autoref{fig:SF6_CF4_He}. 

Similar to the SF$_6$:He mixtures, the CF$_4$:SF$_6$:He mixtures were not able to produce measurable gas gains above a total pressure of 100 Torr. The gain curves which could be measured are seen in \autoref{fig:SF6_CF4_He} (left) and all exhibit exponential behavior and shift to the right with increasing partial pressure of He. Considering the base mixture of CF$_4$:SF$_6$, and  comparing this to the 40 Torr pure SF$_6$ gain curve in \autoref{fig:SF6_Helium_gain}, the amplification fields required to produce a comparable gas gain has been significantly reduced. For example, the amplification field strength required to produce a gas gain > 10$^4$ has reduced from \(\sim\) 28500 V/cm to \(\sim\) 24000 V/cm.



As He is added to the vessel, the maximum attainable gas gain initially rises very slightly with 10 Torr of He. As the partial pressure of He increases to 35 and 60 Torr the maximum gas gain drops. This suggests that the maximum gas gain could drop further as more He is added to the vessel. The maximum gas gains achieved for all CF$_4$:SF$_6$:He mixtures are presented in \autoref{table:CF4SF6He}.

\begin{table}[h]
    \centering
    \captionsetup{justification=centering}
        \caption{Summary of CF$_4$:SF$_6$:He results including maximum stable gas gain, \(G_{max}\), and the minimum/maximum energy resolution, \(ER_{min}\) and \(ER_{max}\).}
        \label{table:CF4SF6He}
        \begin{tabular}{|c c c c|} 
         \hline
         CF$_4$:SF$_6$:He Pressure (Torr) & \(G_{max} \times 10^{4}\) & \(ER_{min}\) & \(ER_{max}\) \\ [0.2ex] 
         \hline
         38:2:0 & 2.61 $\pm$ 0.03 & 0.72 $\pm$ 0.01 & 1.80 $\pm$ 0.04 \\ 
         38:2:10 & 2.99 $\pm$ 0.03 & 0.49 $\pm$ 0.03 &  0.94 $\pm$ 0.03\\ 
         38:2:35 &  2.26 $\pm$ 0.08 & 0.66  $\pm$ 0.02&  1.00 $\pm$ 0.01\\
         38:2:60 & 1.9 $\pm$ 0.1 & 0.86 $\pm$ 0.04 &  2.4 $\pm$ 0.3\\ 

         \hline
        \end{tabular}
\end{table}

Interestingly, the maximum stable gas gains observed in the CF$_4$:SF$_6$:He mixtures is considerably smaller than what was observed in the SF$_6$:He mixtures before sparking ocurred, around \(\sim\) 3 x 10$^4$ compared to \(\sim\) 9 x 10$^4$. This could be due to the lower proportion of SF$_6$ molecules resulting in more of the avalanche electrons being able to propagate freely after the first amplification stage. This could make the probability of electrical breakdown in the second amplification stage more likely. Other possible compounding factors could include photon feedback, due to scintillation in these mixtures, and a reduced attachment cross section due to the larger transfer field strength.

The energy resolution has been evaluated for the various CF$_4$:SF$_6$:He mixtures and can be observed in \autoref{fig:SF6_CF4_He} (right). Additionally the minimum and maximum energy resolutions measured are summarised in \autoref{table:CF4SF6He}. With a small addition of 10 Torr of He the energy resolution appears to initially improve. However, as more He is added to the vessel, the energy resolution deteriorates significantly. Similar to the SF$_6$:He mixtures with larger partial pressures of He, the energy resolution was measured to be > 2 in the 38:2:60 Torr mixture. The worsening energy resolution with increasing partial pressure of He is likely the reason why gas gains could not be measured at 150, 380, and 760 Torr. 


These results demonstrate gas gains on the order of 10$^4$, in not only a low pressure NID gas mixture of CF$_4$:SF$_6$ but also, with additions of He up to a total pressure of 100 Torr. In addition to being a NID gas mixture containing He, these mixtures have the potential to scintillate due to the CF$_4$ component. Further work is required to investigate the scintillating properties of these gas mixtures. 




\section{Conclusions}
\label{sec:conclusions}

In conclusion, the CYGNO collaboration is considering the use of He mixtures, which will extend directional sensitivity to lower WIMP masses. Ideally these mixtures will contain a low partial pressure of fluorine rich gases like CF$_4$ and most importantly the NID gas SF$_6$. A multi-stage Multi-Mesh Thick Gaseous Electron Multiplier (MMThGEM) has recently demonstrated significant charge amplification in the NID gas SF$_6$ at low pressure \cite{McLean2023} and therefore makes a good candidate for testing mixtures containing SF$_6$ and He. 


By first building on a previous optimisation of the MMThGEM, gas gain measurements were taken in 30 and 50 Torr and compared to the previous results in 40 Torr of pure SF$_6$. It was found that 40 Torr of SF$_6$ performed the best by producing the largest gas gain while also producing comparable energy resolutions to the 30 and 50 Torr runs. He was then added incrementally to 40 Torr of SF$_6$ and gas gain measurements were obtained up to a total pressure of 100 Torr. These results demonstrated that maximum gas gains on the order of 10$^4$ are possible in SF$_6$:He mixtures up to at least 100 Torr. As the partial pressure of He increased the maximum stable gas gain began to recede and the energy resolution worsened significantly to > 2 in some cases. The worsening gain and energy resolution is likely the reason why measurements could not be achieved with further additions of He; at total pressures of 150, 380, and 760 Torr. 


In an attempt to find a scintillating NID gas mixture containing He, which could potentially facilitate a complimentary light/charge readout capability, a significant portion of the SF$_6$ base gas was replaced with CF$_4$ in proportions of 38:2 Torr CF$_4$:SF$_6$. A comprehensive optimisation procedure was carried out on the collection and transfer fields in the MMThGEM. Following this, He was added to the mixture gradually. It was found that gas gains on the order of 10$^4$ could be achieved in CF$_4$:SF$_6$:He mixtures up to a total pressure of 100 Torr for the first time. Similar to the SF$_6$:He mixtures, the gain and energy resolution worsened with increasing partial pressure of He and is likely the reason why gain measurements could not be made at 150, 380, and 760 Torr. 

Considering these results, the potential for lower cost atmospheric operation in the future could be realised by minimising the energy resolution via optimisation of the MMThGEM field strengths; improving the energy resolution would also benefit the measurement of the crucial head-tail effect \cite{Thorpe2023}. As it was found that the MMThGEM optimisation could have some dependance on gas composition, an iterative optimisation procedure could be used in future following the initial additions of He. Finally, the light yield of similar CF$_4$:SF$_6$:He mixtures should be demonstrated in the interest of a complimentary light/charge readout method for use in future CYGNO experiments.



\acknowledgments
The authors would like to acknowledge the "University of Sheffield EPSRC Doctoral Training Partnership (DTP) Case Conversion Scholarship" awarded to A.G. McLean.






\end{document}